\documentclass[reprint, amsmath, amssymb, superscriptaddress, aps, prb]{revtex4-2}
\usepackage{float}
\usepackage{graphicx}
\usepackage{microtype}
\usepackage{hyperref}

\begin{document}

\title{Defect Controlled Ferromagnetic Ordering in Au Implanted TiSe$_2$ Nanocrystals}

\author{Utkalika P. Sahoo}%
\affiliation{School of Physical Sciences, National Institute of Science Education and Research, An OCC of Homi Bhabha National Institute, Jatni - 752050, Odisha, India.}%
\author{Spandan Anupam}%
\affiliation{School of Physical Sciences, National Institute of Science Education and Research, An OCC of Homi Bhabha National Institute, Jatni - 752050, Odisha, India.}%
\author{Bidyadhar Das}%
\affiliation{School of Physical Sciences, National Institute of Science Education and Research, An OCC of Homi Bhabha National Institute, Jatni - 752050, Odisha, India.}%
\author{Mrinal K. Sikdar}%
\affiliation{School of Physical Sciences, National Institute of Science Education and Research, An OCC of Homi Bhabha National Institute, Jatni - 752050, Odisha, India.}%
\author{Laxmipriya Nanda}%
\affiliation{School of Physical Sciences, National Institute of Science Education and Research, An OCC of Homi Bhabha National Institute, Jatni - 752050, Odisha, India.}%
\author{Pratap K. Sahoo}%
\email{pratap.sahoo@niser.ac.in}%
\affiliation{School of Physical Sciences, National Institute of Science Education and Research, An OCC of Homi Bhabha National Institute, Jatni - 752050, Odisha, India.}%
\affiliation{Center for Interdisciplinary Sciences (CIS), NISER Bhubaneswar, Jatni - 752050, Odisha, India.}%

\begin{abstract}
   Layered transition metal dichalcogenides (TMDs) are attracting increasing attention because they exhibit unconventional magnetic properties due to crystal imperfections in their usually non-magnetic 2D structure. This work aims to investigate the magnetic response of self-engineered Se deficient TiSe$_2$ thin films, synthesized using chemical vapour deposition. We demonstrate tunability of the ferromagnetic order with the introduction of Au atoms using low energy Au ion implantation, which works as a controlling knob to vary the stoichiometry of Se in TiSe$_{2-x}$. The corresponding isothermal field-magnetization curves fit well with a modified Brillouin J function with J value of 1.5 for Ti$^{3+}$, and 4 for Au$^{3+}$, accounting for the diamagnetism that arises from Au implantation. We propose a qualitative model for the experimentally observed magnetization as a function of ion fluence, corroborated with high-resolution transmission electron microscopy. Depending on the Au nanoparticle size in the implanted samples, magnetization saturates faster at a much lower applied magnetic field than the pristine sample. Our findings hold potential to expand the range of 2D ferromagnetic materials for spintronics and magnetic sensing applications.
   
   \begin{description}
    \item[Keywords] 
    TMDs, CDW, CVD, Ion implantaion, Ferromagnetism, HRTEM.
    \end{description}
\end{abstract}

\maketitle

\section{Introduction}
    In the past decade,  2D transition metal dichalcogenides (TMDs) have been garnering great interest in the research community because of their extensive applications as numerous sensors, transparent flexible nanodevices, and other varied electronic, optoelectronic devices \cite{Guo2019, Baek2017, Pan2020, Sanchez2013, Fiori2014, Jariwala2014, Chen2017, Jiang2020}. They are of the form MX$_2$ (X-M-X), with M being the transition metal (group 3 - 12), one of Ti, Ta, Nb, V; and X being one of the chalcogenides (group 16 elements S, Se, Te). The electronic and magnetic properties of some 2D TMDs can be tuned using anionic substitution \cite{Guo2014}, surface modification \cite{Zhu2016}, and strain engineering \cite{Ma2012}, offering pathways to explore local magnetic moment and spin polarization in these materials \cite{Tong2017, Ahmed2020,Ahmed2019}. Selenides, Sulphides and Tellurides, notably TiS$_2$, TiSe$_2$, NbSe$_2$, NbS$_2$, WSe$_2$ etc. have recently been used in different technological applications such as superconductivity, advanced low-power electronics, voltage-controlled oscillators, ultra-fast electronics, electrochemical devices, etc. \cite{Zhang2014, Wei2017, Goli2012, Monney2015, Ozaydin2015}. Kaur and Eda et al reported that the electronic properties of these materials have been tuned between semiconducting, semi-metallic, metallic, and insulating depending on their phases, polytypes and symmetry \cite{Kaur2017, Eda2011}. Cai et al \cite{Cai2015} reported a phase-incorporation strategy by dual native defects in the TMD to induce ferromagnetism into the originally non-magnetic MoS$_2$ nanosheets, arising from the exchange interactions within the vacancy site and the dangling 4d Mo$^{4+}$ centres.
    
    Titanium Diselenide (TiSe$_2$), with its small semiconducting bandgap, is one of the most explored in the TMD family. It has been shown to be an excellent system for observing the low-temperature Charge Density Wave (CDW) state \cite{Salvo1976}, and it has been in the spotlight ever since. Both bulk and thin-film TiSe$_2$ show the CDW transition at around 205 K while forming a superstructure, which is attributed to a periodic lattice distortion \cite{Rossnagel2011, Holt2001}. The CDW order can be tuned after the introduction of disorder via intercalation, pressurization, straining, electron beam irradiation, electrical bias or doping with a guest element; and reducing the thickness and dimensionality of the thin film \cite{Xi2015, Joe2014, Calandra2015}. The introduction of the disorder can alter the basic properties of the material and make it a wideband semiconductor too, which might be a gateway for discovering high-temperature CDW states, paving the path for high-temperature CDW electronic devices. TiSe$_2$ crystal belongs to the TMD family in the $p\overline{3}m1$ (164) space group, crystallizing in hexagonal and triangular form. The crystal structure of TiSe$_2$ consists of an octahedrally co-ordinated Ti$^{4+}$ atom at the centre of the complex, sandwiched between layers of Se$^{2-}$ atoms. This makes Ti-Se-Ti layers with a small Van der Waals (VdW) gap of ($\sim0.61$nm) along the c axis and strong intralayer covalent bonds along the ab plane \cite{Li2019, Yan2018}. It is known that the covalent Ti-Se interactions in TiSe$_2$ quench magnetic moments, rendering it non-magnetic. Recently though, magnetism and spin polarization in TiSe$_2$ was introduced through methods like magnetic atom intercalation as shown by Luo, Morosan et al \cite{Luo2015, Morosan2006}, and through vacancies, as shown by Tong et al \cite{Tong2017}. Tong et al introduced Se vacancy by doping Ti atoms in TiSe$_2$, creating Ti$^{3+}$ through the transfer of a single electron to 3d orbital, eventually leading to ferromagnetic behaviour in the material. 
    
   In this work, we have synthesized TiSe$_{2-x}$ nanocrystals with intrinsic Se defects. We have been able to gain precise control over the defect density with the help of low energy Au ion beam irradiation, which lets us exercise the freedom to choose the kind of magnetic ordering. As the concentration of Au increases, the diamagnetic property of the sample increases systematically. However, we note that the saturation magnetization bumps up a notch at lower fluences and decreases as we go forward. The evolution of Au nanoparticles to nanoclusters with Au ion fluence ($\phi$) in TiSe$_{2-x}$ was experimentally observed and modelled to explain the magnetic ordering. Though Tong et al \cite{Tong2017} observed Se deficient induced magnetism in TiSe$_2$, they had no actual control over the defect density. This work aims to explore defect controlled magnetism while establishing a possible explanation behind the tunability.
   
\section{Experimental}
\begin{figure}[H]
        \centering
        \includegraphics[width=1\columnwidth]{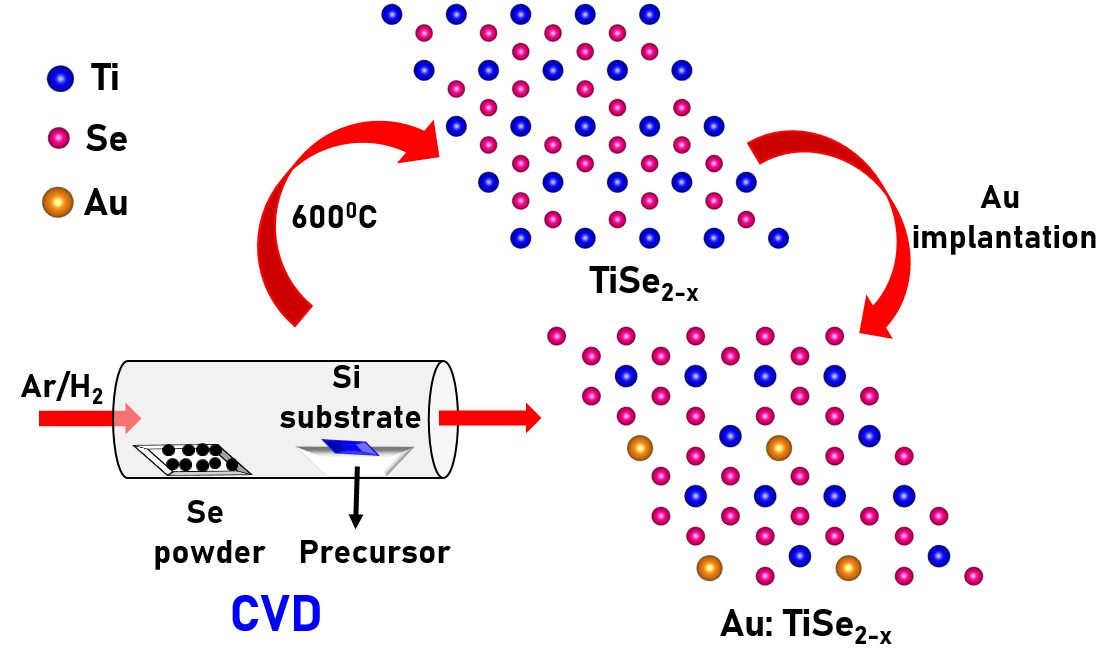}
        \caption{Schematic of the CVD system
        for the TiSe$_{2-x}$ thin film and after Au ion implantated Au:TiSe$_{2-x}$ thin film.}
        \label{cvd}
    \end{figure}

    TiSe$_{2-x}$ nanocrystals with controlled Ti and Se raw ratio were grown on the Si substrates using a single zone Chemical Vapour Deposition (CVD) system in ambient pressure. Consistent and repeatable growth of hexagonal TiSe$_{2-x}$ nanocrystals on Si substrate was achieved using a modified version of the synthesis method provided by Wang et al \cite{Wang2018}. The TiO$_x$ and LiCl mixture was kept at the centre of the CVD tube and maintained at 600$^\circ$C throughout the reaction. Se powder was kept upstream at a temperature of about 260$^\circ$C. Forming gas (mixture of H$_2$ (5$\%$) and Ar (95$\%$)) was passed through the tube at 100 sccm. After synthesis, the samples were irradiated with 20 KeV Au ions at fluences of $1\times10^{15}$, $2\times10^{15}$, $1\times10^{16}$, and $2\times10^{16}$ ions cm$^{-2}$. The schematic of the setup along with the formation of TiSe$_{2-x}$ and Au defects in the matrix was illustrated in Fig. \ref{cvd}.
   
    The phase identification of TiSe$_{2-x}$ samples, before and after Au ion implantation was carried out by Rigaku X-ray diffraction (XRD) using Cu-K$\alpha$ radiation with a wavelength of $\lambda$ = 1.5406\AA. The surface morphology of as-deposited and Au-irradiated samples were characterized using Field Emission Scanning Electron Microscopy (FESEM) (Sigma-ZEISS). The elemental distribution of Ti, Se and Au was confirmed using an Energy Dispersive X-Ray Spectroscopy (EDXS) incorporated into the FESEM. We expect the formation of non functionalized Au nanoparticles in TiSe$_{2-x}$ with increasing fluence, which was later confirmed and visualized using High-Resolution Transmission Electron Microscopy (HR-TEM) micrographs with field emission electron source at 200 keV (Jeol-F200). Superconducting Quantum Interference Device (SQUID) magnetometry was used for magnetization measurement at all the temperatures ranging 5 K - 300 K.
 
\section{Results and Discussion}
    Grazing Incidence X-Ray Diffraction (GIXRD) measurements were performed to obtain information on TiSe$_{2}$ formation, and track any changes after Au ion implantation.
      \begin{figure}[H]
        \centering
        \includegraphics[width=1\columnwidth]{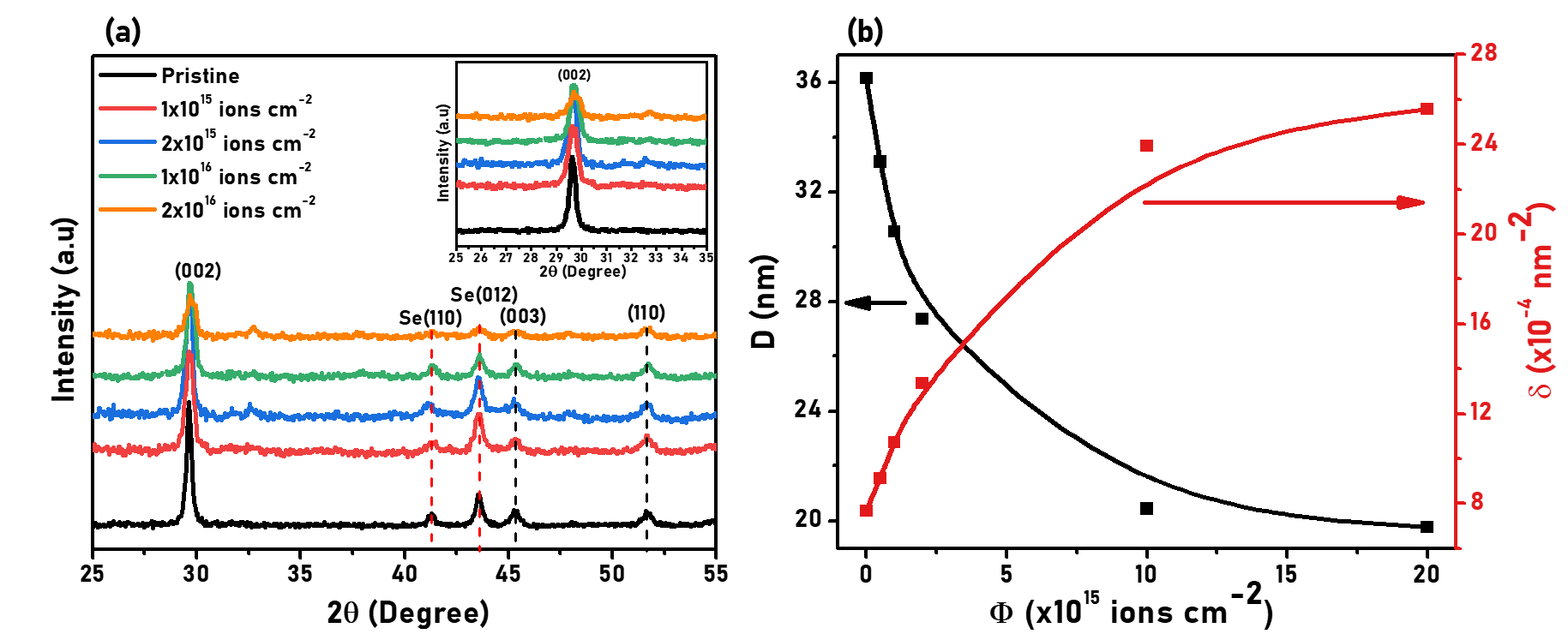}
        \caption{(a) XRD pattern of as deposited and Au implanted TiSe$_2$ samples. Inset shows reduction in intensity and broadening of the prominent peak of TiSe$_2$ (002) plane after Au ion irradiation. (b) The Debeye-Scherrer calculated crystallite size and dislocation density as a function of fluence.}
        \label{xrd}
    \end{figure}
    
    Figure \ref{xrd}(a) shows the XRD pattern with peaks of (002), (003), and (110) planes as compared with the JCPDS file number 651885, within the 2$\theta$ values of 25 to 55$^{\circ}$ corresponding to the crystal's p$\bar{3}$m1 TiSe$_2$ phase. The peaks at 2$\theta$ values 41.35 and 43.6 correspond to the (110) and (012) planes of the residual Se on the substrates during CVD growth.  All the peaks seem prominent, dominated by the (001) orientation; which confirms the formation of highly crystalline and oriented nanocrystals. 
    
    The calculated lattice parameters come out to be  $a = b = 3.533$ {\AA}, and $c = 6.010$ {\AA} using the relation \cite{Bindu2014} for hexagonal TiSe$_2$ crystal structure, which is in the ballpark for bulk TiSe$_2$ \cite{Behera2019}. The parameter ``c" shows a slight decline with the fluence, however, the a and b values change quite significantly. This could be ascribed to the compressive strain along the a and b direction due to all the Se vacancy sites. All the TiSe$_2$ peaks broaden as fluence is increased indicating the degradation of crystallinity. The inset in Fig. \ref{xrd}(a) shows the magnified version of the most prominent (002) peak. Crystalline size (D) and the dislocation density ($\delta$) were calculated using the Debye-Scherrer equation and the relation ($\delta$=1/D$^2$) \cite{Kumar2017}  as a function of ion fluence, shown in Fig. \ref{xrd}(b). As expected, the mean crystalline size decreases from 36 nm to 18 nm as the ion fluence increases up to $2\times10^{16}$ ions cm$^{-2}$. We also see that the $\delta$ value increases from $6\times10^{-4}$ nm$^{-2}$ to $26\times10^{-4}$ nm$^{-2}$ as the ion fluence increases, shown in Fig. \ref{xrd}(b). 
    
\begin{figure}[H]
        \centering
        \includegraphics[width=1\columnwidth]{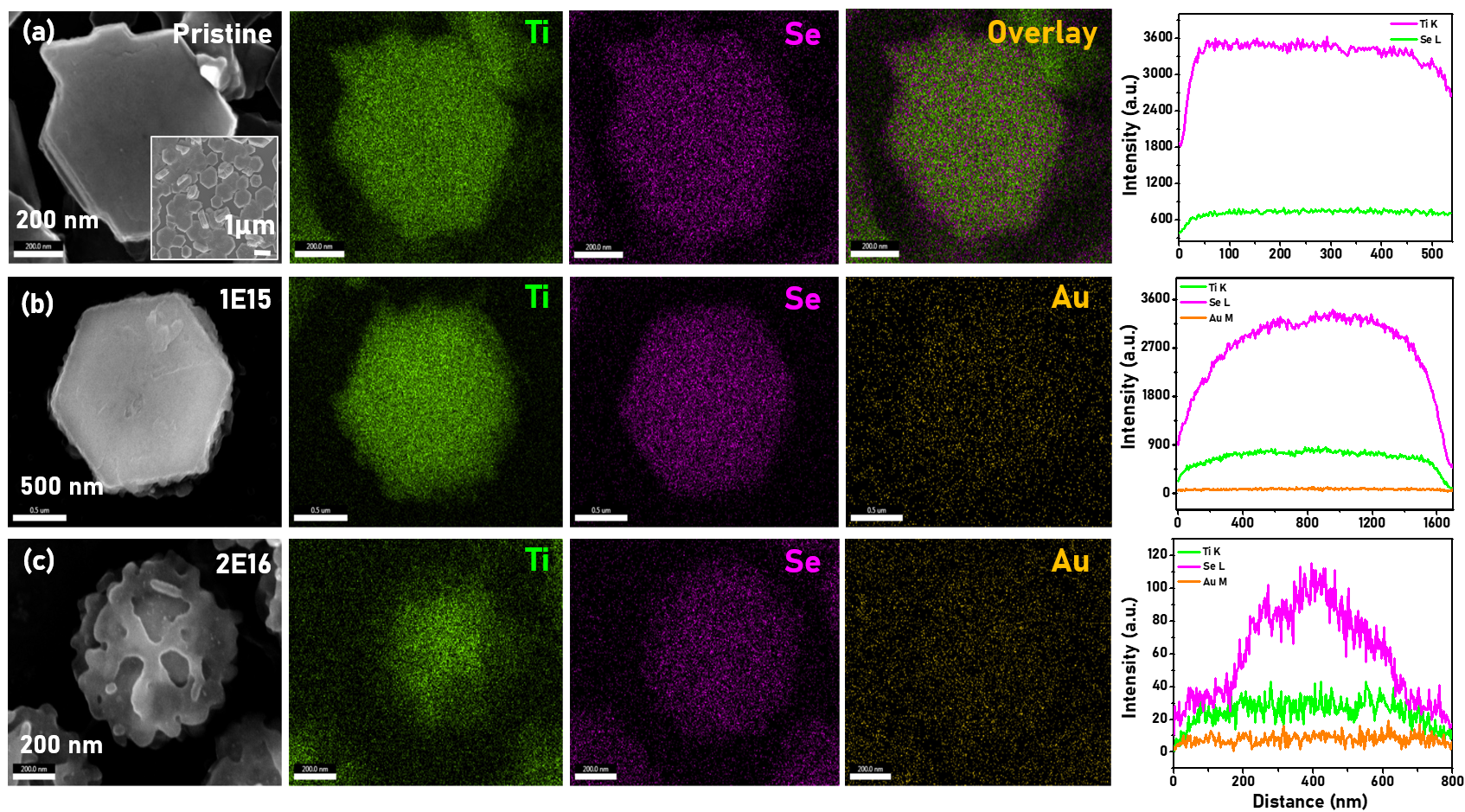}
        \caption{Mapped SEM image of pristine and irradiated sample with fluence range of  1$\times10^{15}$- 2$\times10^{16}$ ions cm$^{-2}$.}
        \label{mappedsem}
    \end{figure}

    The surface morphology of the pristine and Au-irradiated TiSe$_{2-x}$ was studied using FESEM and EDXS. In FESEM images, we see the formation of hexagonal crystals, with length varying from 1 - 2 $\mu$m as shown in Fig. \ref{mappedsem}. From cross-sectional FESEM, the thicknesses of the crystals is found to be in the range of $\approx$ 150 - 200 nm. The crystals progressively deform and become porous with the increase in ion fluences. A representative FESEM and EDXS mapping of fluence $1\times10^{15}$  and $2\times10^{16}$ ions cm$^{-2}$ is shown in Fig. \ref{mappedsem}(b) and (c). 
    From elemental mapping, the distribution of Ti and Se mimic the hexagonal shape in all the samples, whereas Au is all over the surface area. The line scans demonstrate the uneven distribution of Ti, Se and Au after ion implantation. The Ti:Se ratio for the as-grown and implanted samples at different fluences are tabulated in Table \ref{tab:edx}. The calculated ratio of Ti/Se decreases as we go to higher fluences, which matches with our sputtering yield calculation using TRIDYN \cite{Mller1984} and SRIM \cite{Ziegler2010}. We see that the Ti:Se ratio systematically decreases with the increase in ion fluence, with the pristine standing at 1:1.80, decreasing to 1:1.73 and then to 1:1.09 as soon as the sample gets irradiated. 
    \begin{table}[H]
        \centering
        \resizebox{0.9\columnwidth}{!}{%
        \begin{tabular}{|c|c|c|c|c|c|}
        \hline
        Element & Pristine & 1$\times10^{15}$ & 2$\times10^{15}$ & 1$\times10^{16}$ & 2$\times10^{16}$ \\ \hline
         Ti:Se & 1:1.8 & 1:1.7 & 1:1.6 & 1:1.2 & 1:1.1 \\ \hline
        Au & - & 0.9 & 1.8 & 9.1 & 9.6 \\ \hline
        \end{tabular}%
        }
        \caption{Elemental composition of the pristine and Au ion irradited samples from FESEM-EDXS analysis.}
        \label{tab:edx}
    \end{table}
    This is caused due to sputtering of Ti and Se ions as the high energy Au atoms strike the lattice at normal incidence. The atomic percentage of Au ions increases with fluence which is shown in Table 1. It is well known that the nuclear energy loss ($2.328\times10^3$ eV/nm) of 20 keV Au ions in TiSe$_2$ yields a large amount of sputtering of Ti and Se atoms. The penetration depth of 20 keV Au is $\approx$ 20 nm (SRIM), extending till a depth of 30 nm, due to dynamic sputtering during ion implantation. The total calculated sputtering yield of Ti and Se are $\approx$ 6 atoms/ion, which is quite high to make TiSe$_{2-x}$ porous at high fluence.
    \begin{figure}[H]
        \centering
        \includegraphics[width=1\columnwidth]{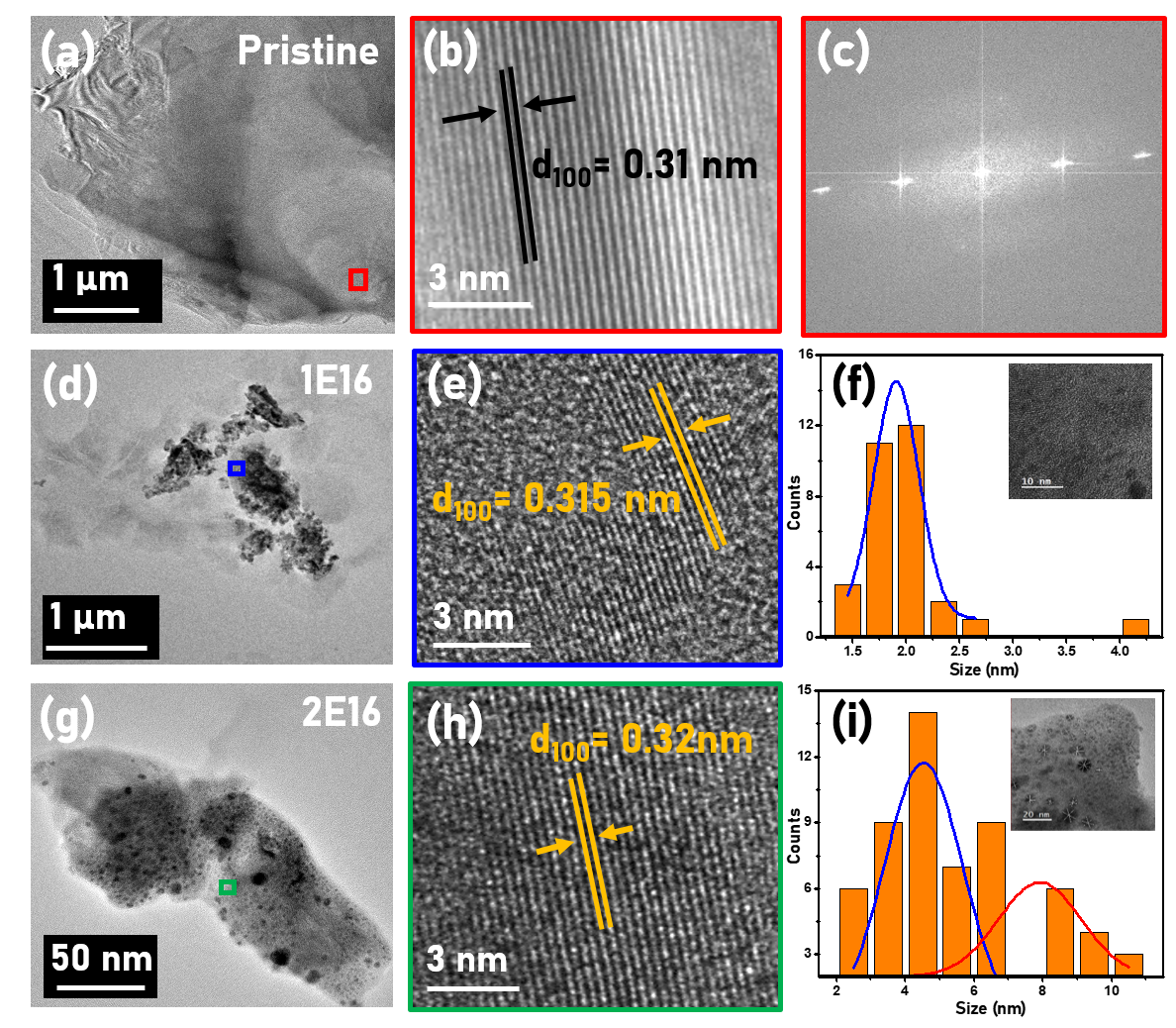}
    \caption{(a) Large area TEM image of the as prepared TiSe$_{2-x}$ sample. (b) Zoomed in view of the sample showing its (100) plane. (c) SAED pattern of the large area TiSe$_{2-x}$ (d)and (g) Large area TEM image of Au:TiSe$_{2-x}$ sample at a fluence of 1$\times10^{16}$  and 2$\times10^{16}$ ions cm$^{-2}$. (e) and (h) Zoomed in view of the irradiated sample with marked Au (111) patches having d = 0.24nm (f) and (i) Average domain size of Au nanoparticles, determined as 2 nm and 5.81 nm respectively.}
        \label{tem}
    \end{figure}
    To characterize the effect of Au ion implantation on TiSe$_{2-x}$ nanocrystals, HRTEM analysis was performed. HRTEM images for the pristine and the samples irradiated at 1$\times10^{16}$  and 2$\times10^{16}$ ions cm$^{-2}$ are shown in Fig. \ref{tem}. For the TEM samples, some flakes of TiSe$_{2-x}$ were transferred from the Si surface onto the Cu TEM grid. Figure \ref{tem}(a-c) shows the TEM micrograph, high-resolution fringe patterns (zoomed view of the red squared region), and corresponding Selected Area Diffraction Pattern (SAED) pattern of the pristine sample. It was observed that the fringes are oriented along the (100) plane and the calculated d value is 0.31 nm shown in Fig. \ref{tem}(b). The Au ion implantation introduced defects and nanoscale amorphous regions and nanoparticles inside the TiSe$_{2-x}$ matrix. Structural distortion is visible in the 1$\times10^{16}$ ions/cm$^2$ irradiated sample, with a distinct increase in the lattice spacing to 0.315 nm (Fig. \ref{tem}(e)) alongside the formation of small size Au nanoparticles as shown in the inset of Fig. \ref{tem}(f). The particle size was calculated and the average domain size was determined to be 2 nm shown in Fig. \ref{tem}(f). At the highest fluence of 2$\times10^{16}$ ions cm$^{-2}$ the lattice spacing increases to 0.32 nm (Fig. \ref{tem}(h)) because of stress developed due to larger Se deficient in TiSe$_{2-x}$.  Due to heavy ion irradiation-induced self-heating, many small units agglomerate and tend to form larger Au nanoparticles as shown in the inset of Fig. \ref{tem}(i). They show up as dark patches with nanoparticle sizes varying from 2 - 12 nm with a bimodal distribution.  Two Gaussian fittings centred around 5 nm and 8 nm are the average size of the Au nanoparticles shown in Fig. \ref{tem}(i).

    Ferromagnetic behaviour in TiSe$_2$ is especially astounding because the Ti atoms are hybridized with Se atoms by covalent Ti–Se interaction, which quenches the magnetic moment keeping it non-magnetic. Most of the reports on ferromagnetism in TiSe$_2$ have been studied through several ways: (a) Via introducing magnetic atoms leading to magnetism or spin polarization \cite{Luo2015}, (b) By way of incorporating single Se-anion defects and intercalating Ti atoms inside the TiSe$_{1.8}$ material \cite{Tong2017}, (c) By using secondary phases \cite{Prusty2016}. Tong et al \cite{Tong2017} demonstrated that TiSe$_{1.8}$  is ferromagnetic with local magnetic moments and spin polarization due to single Se-cation and Ti intercalation. But in our case, the pristine sample has intrinsic Se defects, showing up in the form of TiSe$_{2-x}$, which varies with Au incorporation. So exploring the tunable magnetic properties by an effective strategy to induce and control the ferromagnetism is highly desirable. The incorporation of Au atoms by low energy ion implantation was used as a regulating knob to control the Se vacancies, Au intercalation and Ti intercalation in order to trigger the magnetic ordering. 
    \begin{figure}[H]
        \centering
        \includegraphics[width=1\columnwidth]{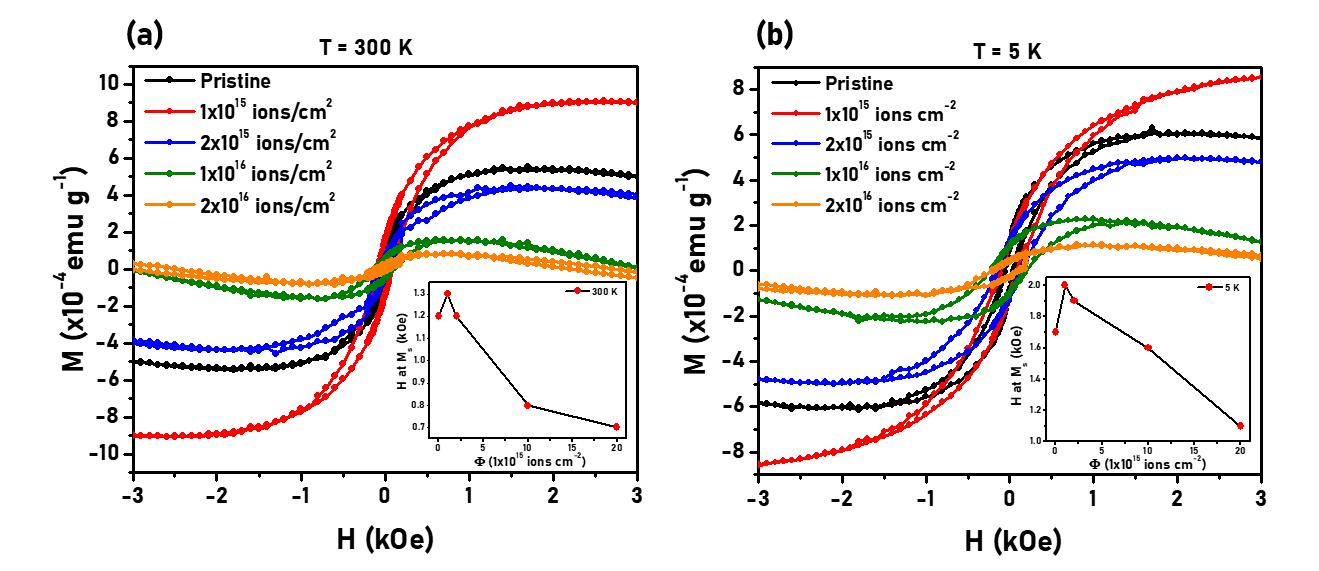}
        \caption{ (a) and (b) Show the hysteresis curve of pristine and post irradiated samples at temperatures of 300 K and 5 K. Inset shows applied magnetic field (B) at saturation magnetisation (M$_S$) of pristine and irradiated samples at 300 K and 5 K.}
        \label{mhcombo}
    \end{figure}
    Figure \ref{mhcombo}(a) and (b) show the ferromagnetic response of the as-deposited and the post irradiated samples at 300 K and 5 K, respectively with different ion fluences in the range of 1$\times10^{15}$- 2$\times10^{16}$ ions cm$^{-2}$.  It was noticed that the as-prepared TiSe$_{1.8}$ system illustrates a characteristic ferromagnetic S-shaped (M-H) curve. In contrast, diamagnetic behaviour starts showing up with the increase in ion fluence.  As the temperature decreases from 300 K to 5 K, the width of the hysteresis loop becomes more prominent for all samples. In all temperature ranges, we observed a sharp increase in the saturation magnetisation at the lowest fluence and a steady decrease from thereon. The magnetic moment saturates faster with increasing ion fluence. The insets of Fig. \ref{mhcombo} (a) and (b) show the behaviours of the applied magnetic field for M$_s$ as a function of ion fluence. Both at 300 K and 5 K the magnetic field at M$_s$ increases for the lowest fluence, then decreases as a function of ion fluence. A similar trend was observed for all the temperatures. During ion implantation, both Se and Ti atoms get sputtered out, which creates more Se vacancies, Ti displacements, and Au incorporation. Such synergistic effect creates controllable defect states, eventually reducing the total magnetic moment per unit volume, requiring less applied magnetic field for saturation magnetization at higher fluence.
    \begin{figure}[H]
        \centering
        \includegraphics[width=1\columnwidth]{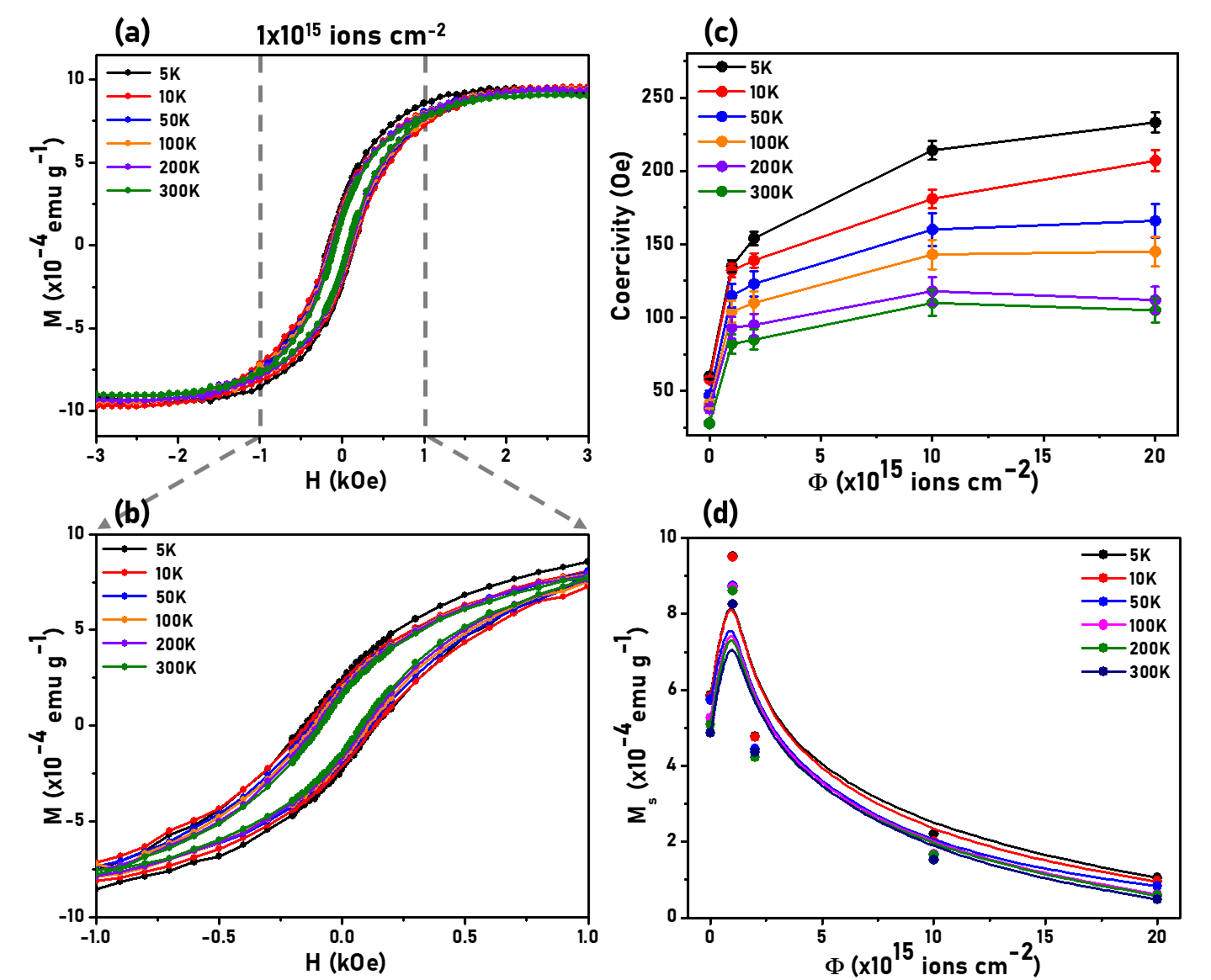}
        \caption{(a) Hysteresis curve at fluence of 1$\times10^{15}$ ions cm$^{-2}$ at variable temperature. (b) Shows the zoomed view of (a) at an applied magnetic field of -1000 Oe to 1000 Oe. (c) and (d) Show the coercivity and saturation magnetisation of pristine and post irradiated samples at different temperatures. }
        \label{mh}
    \end{figure}
    
    A representative plot with a complete systematic study of the S-shaped hysteresis loop in a temperature range of 5 K to 300 K at a fluence of 1$\times10^{15}$ ions cm$^{-2}$ is depicted in Fig. \ref{mh}(a). Magnified view of the relevant region in the applied magnetic field of -1000 Oe to 1000 Oe is shown in Fig. \ref{mh}(b). With the increase in temperature, the hysteresis width decreases, resulting in a decrease in its coercivity. The coercivity and saturation magnetisation (M$_s$) values were extracted from the M-H curve for as-deposited and post irradiated samples for all the variable temperatures and displayed in Fig. \ref{mh}(c) and (d). The coercivity values of the samples increase with ion fluence at all the temperatures as shown in Fig. \ref{mh}(c). The maximum change of coercivity observed is from 45 Oe to 235 Oe at 5 K, whereas the change is from 40 to 75 Oe at 300 K. In Fig. \ref{mh}(d), we have observed that the M$_s$ value seems to bump up at a fluence of 1$\times10^{15}$ ions cm$^{-2}$ and then decrease as we increase the ion fluence.
   
\begin{figure}[H]
        \centering
        \includegraphics[width=1\columnwidth]{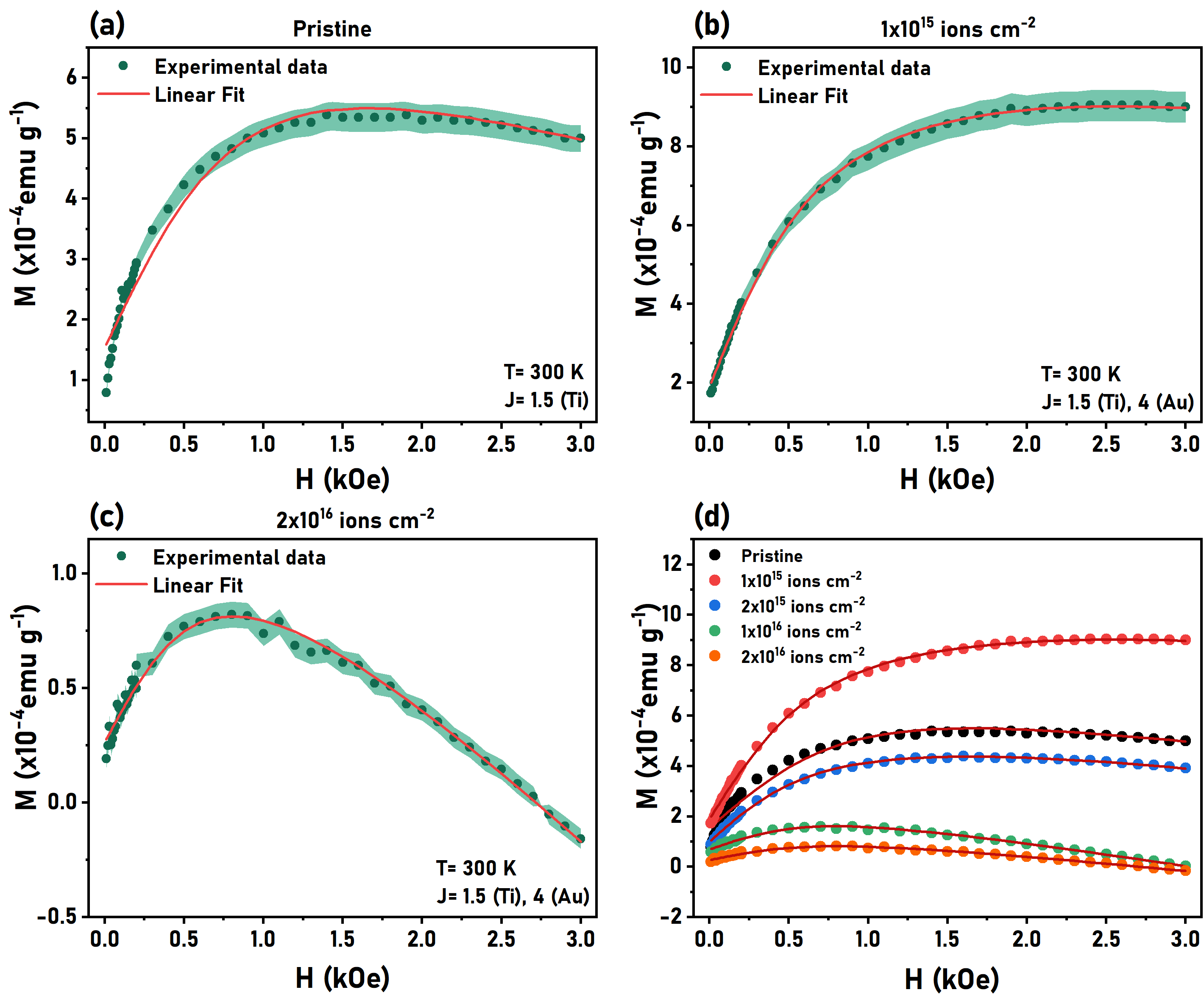}
        \caption{(a-c) Show the fits, and experimental data for M-H curves at room temperature, at different ion fluences. (d) Shows the combined fits and plots for M-H curves at all ion fluences}
        \label{mhcombobig}
    \end{figure}

    It is observed from the hysteresis loop of Fig. \ref{mhcombo}(a) and (b) that the diamagnetic contribution becomes higher with the decrease in saturation magnetisation. In this work, we show that competing contributions between ferromagnetic (FM) and diamagnetic (DM) phases result in opposite effects: enhancement of coercivity and decrease in the saturation magnetisation. The source of FM and diamagnetic dip can be explained by all the different oxidation states from the different atoms present in the matrix. Since we are bombarding Au atoms on Se deficient samples, we expect the presence of mixed oxidation states from Titanium, which is Ti$^{3+}$,  Ti$^{4+}$\cite{Tong2017}, and Au$^{3+}$.  To explain the coexistence of FM and DM phases, we resort to the modified Brillouin function \cite{Saddik2020}. The positive arms of the M-H curves were fitted using the Brillouin J relation to calculate the magnetization theoretically using the $J, L, S$ triplets of atoms contributing to magnetization. The following Brillouin function was used which takes into account the contribution of both FM and DM phases.
    \begin{eqnarray}
        M = M^{1}_{s} B_J^{(1)}(B) + M^{2}_{s} B_J^{(2)}(B) + \chi_{corr} B + const.
    \end{eqnarray}
    Where M$_{sat}$ is the saturation magnetization, and the Brillouin J function is defined with $L_n, S_n, J_n$ as:
    \begin{eqnarray}
        B_J^{(n)} (B) = \frac{2J_n + 1}{2J_n}\coth&\Biggl(\frac{2J_n + 1}{2J_ n}x(B)\Biggr)\\ 
        &- \frac{1}{2J_n}\coth\Biggl(\frac{1}{2J_n}x(B)\Biggr)
    \end{eqnarray}
    where
    \begin{eqnarray}
        x &= J\frac{g \mu_B B}{k_B T}\\
        B &= \mu_r \mu_o H
    \end{eqnarray}
    and $g$ is defined as:
    \begin{equation}
        g = 1 + \frac{J_n(J_n+1) + S_n(S_n+1) - L_n(L_n+1)}{2J_n(J_n+1)}
    \end{equation}
    $\chi_{corr}$ represents the effective magnetic susceptibility; its value is negative for DM contributions and positive in case of a FM one. It should be noted that the linear correction would correct for diamagnetic tendencies in the sample, which is observed at the farther edge of the hysteresis curves \cite{Saddik2020} in Fig. \ref{mhcombobig}.

    The fitting of Brillouin J function with the $J, L, S$ values corresponding to Ti$^{3+}$'s $^2D_{3/2}$ of the as-grown sample without any diamagnetic correction term is shown in Fig. \ref{mhcombobig}(a), as the M-H hysteresis curve shows ferromagnetic behaviour and has no prior reports suggesting diamagnetism in pristine systems.  Tong et al \cite{Tong2017} demonstrate that the intrinsic local magnetic moments arise from Ti$^{3+}$'s 3d$^1$ electron configuration in TiSe$_{1.8}$.  Figure \ref{mhcombobig}(b) and (c) show the fits for the Au implanted samples at a fluence of 1$\times10^{15}$ ions cm$^{-2}$ and 2$\times10^{16}$ ions cm$^{-2}$ respectively. The fits seem to stay within the experimental error (filled area) in the measurement of magnetization for all the samples shown in Fig. \ref{mhcombobig}(a-c). The two parts of the Brillouin function correspond to the contribution by Au$^{3+}$'s $^4F_{3}$ state along with Ti$^{3+}$'s $^2D_{3/2}$ with a diamagnetic correction term stemming from the Au implantation \cite{Prusty2016, Hori2004, Li2011}. We don't fit for the Au's $+1$ state here, because it corresponds to d$^0$, which doesn't give us a definite $J, L, S$ triplet.  One can see that the diamagnetic contribution increases as the fluence increases (Fig. \ref{mhcombobig}(c)). 

    \begin{figure}[H]
        \centering
        \includegraphics[width=1\columnwidth]{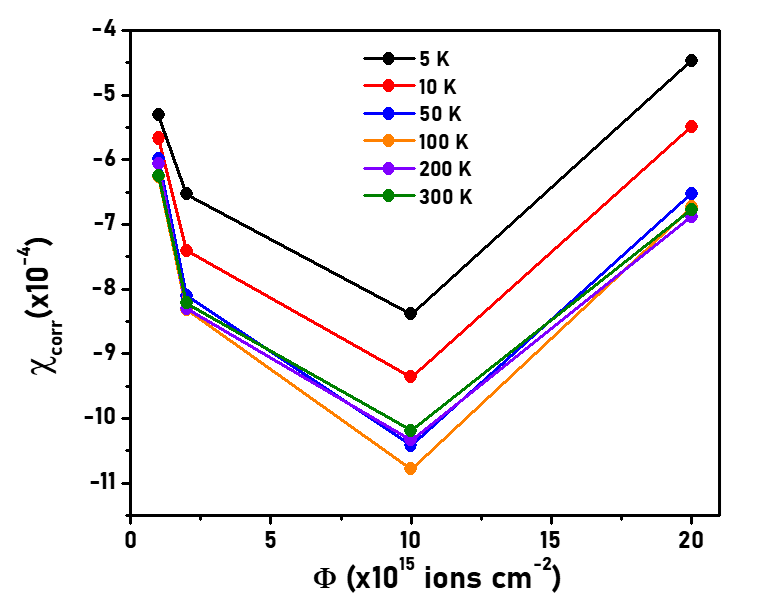}
        \caption{(a) Shows the calculated magnetic susceptibility ($\chi_{corr}$) at all variable temprerature from 5 K - 300 K as a function of ion fluence}
        \label{xcorr}
    \end{figure}
    
    Figure \ref{xcorr} shows the correction ($\chi_{corr}$) factor for the Au implanted samples at different fluences calculated from the fits. The value of $\chi_{corr}$ is negative due to the diamagnetic behaviour, with the magnitude increasing systematically till 1$\times10^{16}$ ions cm$^{-2}$, then dropping sharply at the highest fluence of 2$\times10^{16}$ ions cm$^{-2}$, which is attributed as the formation of a high number of non-magnetic Au agglomerates.

\begin{figure}[H]
        \centering
        \includegraphics[width=1\columnwidth]{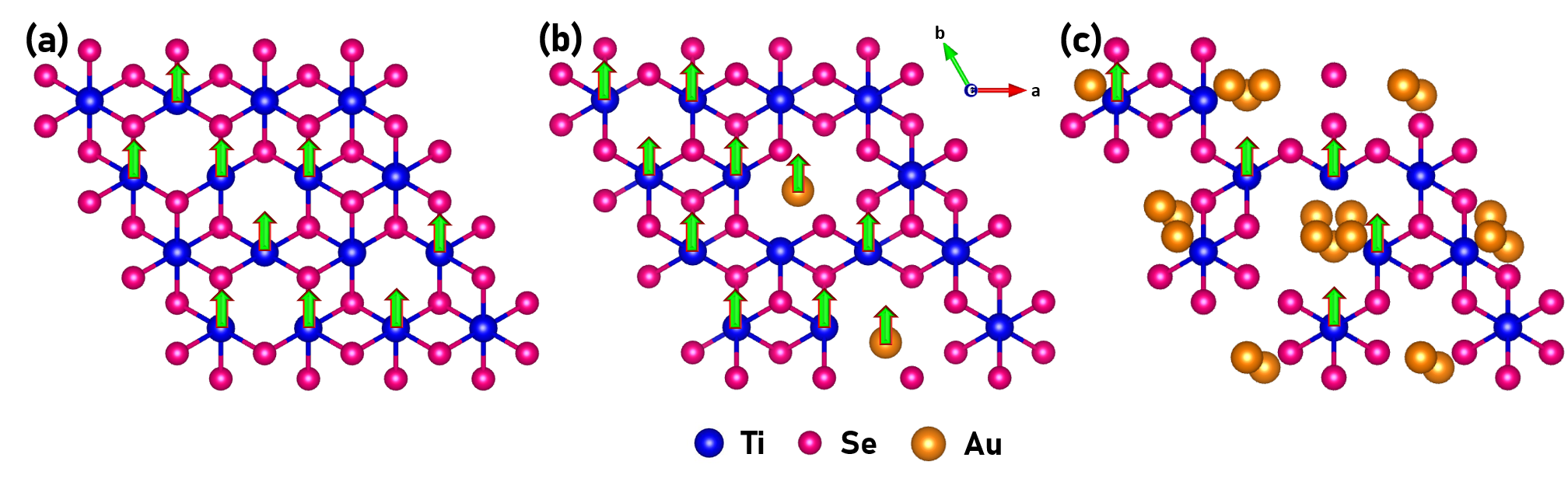}
        \caption{Model explaining origin of magnetisation in (a) Pristine samples, mostly from the Ti atoms around the Se deficient centers (b) Lower fluences, originating from Au nanoparticles and deficient Ti centers and (c) Higher fluences suffering from porous lattices and non-magnetic Au clusters }
         \label{model}
    \end{figure}

    It was observed that at all the temperatures, M$_s$ shows a sharp increase at lower fluences, and then a systematic decline as the fluence goes higher. To explain such an effect, we have modelled the modification of the TiSe$_2$ matrix by Au ion incorporation as illustrated in Fig. \ref{model}.  It is very clear that the Ti$^{3+}$ states created due to the Se deficiency is the source of magnetization in case of the pristine TiSe$_{1.8}$ sample \cite{Tong2017, Xiao2019}.  Figure \ref{model} (a) shows a model for Se deficient pristine samples where the Ti$^{3+}$ states have a net magnetic moment. We have explored an approach to effectively control the magnetization by introducing Au ions, which can prompt magnetization in non-magnetic materials depending on the nanoparticle size. Li et al \cite{Li2011} have shown us that the magnetic properties of Au nanoparticles are most pronounced when they are at around 2 - 4 nm in size. At lower fluences, the Au atoms are randomly distributed. At the same time, some of the Se and Ti atoms are knocked out due to ion beam sputtering. The calculated sputtering yield of Ti and Se at a fluence of $1\times10^{15}$ ions cm$^{-2}$ is 1.7 and 4.3 atoms/ion respectively. Since the sputtering yield of Ti is lesser as compared to Se, the probability of Ti$^{3+}$ formation is high because the excess Se vacancies would transfer one electron to the empty 3d orbital of Ti atoms. At the same time, Au atoms are distributed randomly all over the area of irradiation, creating Au nanoparticles in the matrix. Figure \ref{model}(b) shows the model for Ti$^{3+}$ and distributed Au$^{3+}$ sample at a lower fluence. HRTEM micrographs with the Au nanoparticle distribution in Fig. \ref{tem}(f) seem to agree with our model. These combined factors provide us with a good explanation for the higher saturation magnetization at a fluence of 1$\times10^{15}$ ions cm$^{-2}$ compared to the as grown sample.
   
    A decrease in M$_s$ was observed with an increase in ion fluences. This may be due to the decrease in the net volume concentration of atoms contributing to magnetization.  As the fluence increases the sputtering yield goes up for Ti atoms and down for Se atoms. At higher fluences (2$\times10^{16}$ ions cm$^{-2}$), the sputtering yield is 1.9 and 4.2 atoms/ion for Ti and Se respectively. This yield increases porosity in TiSe$_{1.8}$ thin films as shown in Fig. \ref{mappedsem}(c) in the FESEM image, which is the major reason for the decline in magnetization. At this fluence, M$_s$ goes down to a minimum because (a) Volume concentration of  Ti$^{3+}$ decreases due to sputtering out of a large fraction of Ti atoms, (b) Au ions agglomerate into nanoclusters with an average size of 4 - 8 nm (as shown in Fig. \ref{tem}(i) of HRTEM image), which increase the DM contribution and decrease the magnetization. Figure \ref{model}(c) illustrates the clustering stage of the sample. The systematic increase in the diamagnetic behaviour stems from Au nanoparticles sitting within the lattice.  It is clear that by engineering defects via precise Au ion incorporation, one can indeed control ferromagnetism in TiSe$_2$.

\section{Conclusion}
   We have demonstrated that the ferromagnetic order can be tuned in CVD grown 2D TiSe$_{1.8}$ TMDs by 20 keV Au ion implantation. The structural changes and stoichiometric variations of Ti and Se after Au incorporation were confirmed from XRD, FESEM with EDXS, and HRTEM. The intrinsic ferromagnetism in TiSe$_{1.8}$ arises from the Ti$^{3+}$ oxidation state generated by Se vacancies. Depending on the ion fluence, the system became disordered by the introduction of Au nanoparticles and nanoclusters. When the Au nanoparticle size remained within 2 - 4 nm, the system showed ferromagnetism dominantly. Larger Au nanoparticles increased the diamagnetic contribution.  At the highest fluence, TiSe$_{1.8}$  matrix turned porous with a high dislocation density, and Au nanoclusters went  on to reduce the FM contribution and enhance the DM contribution. The DM contributions ($\chi_{corr}$) were estimated by fitting the hysteresis curve using a modified Brillouin function incorporating both the Ti$^{3+}$ and Au$^{3+}$ oxidation states with J values of 1.5 and 4, respectively. Precise Au incorporation with a balanced FM and DM contribution can pave a new way to tune layered TMD materials' magnetic properties and shine new light on their potential applications in spintronic and magnetic sensing devices.
  
\section*{Acknowledgments}
    This work was supported by National Institute of Science Education and Research, DAE, India. The authors acknowledge the staff of low energy ion beam facility of Institute of Physics (IOP), Bhubaneswar for providing stable beams during ion implantation. PKS thanks Dr. Kartik Senapati for  fruitful scientific discussion.  

\bibliography{output}
\end{document}